\documentclass[a4paper,11pt]{article}
\pdfoutput=1
\usepackage{jheppub} 

\usepackage{graphicx, epsfig}
\usepackage{color}
\usepackage{mathrsfs}
\usepackage {amssymb}
\usepackage {amsmath}
\usepackage{float}

\newcommand{\be}{\begin{equation}}
\newcommand{\ee}{\end{equation}}
\newcommand{\beq}{\begin{eqnarray}}
\newcommand{\eeq}{\end{eqnarray}}

\newcommand{\bea}{\begin{eqnarray}}
\newcommand{\eea}{\end{eqnarray}}
\newcommand{\nn}{\nonumber}
\newcommand{\bl}{U(1)_{\mathrm B-L}}

\newcommand{\Tr}{{\rm Tr}}

\begin{document}

\begin{flushleft}
KCL-PH-TH/2013-17,  LYCEN 2013-05 \\
\end{flushleft}

\title{\boldmath Is F-term hybrid inflation natural within minimal supersymmetric SO(10)?}
\vspace{1cm}
\author[a]{Giacomo Cacciapaglia,}
\author[b]{Mairi Sakellariadou}

\affiliation[a]{Universit\'e de Lyon, F-69622 Lyon, France; Universit\'e Lyon 1, Villeurbanne;\\
CNRS/IN2P3, UMR5822, Institut de Physique Nucl\'eaire de Lyon
F-69622 Villeurbanne Cedex, France} 
\affiliation[b]{King's College London, Department of Physics, Strand,
London WC2R 2LS, UK}

\emailAdd{g.cacciapaglia@ipnl.in2p3.fr}
\emailAdd{mairi.sakellariadou@kcl.ac.uk}

\abstract
{We examine whether F-term supersymmetric hybrid inflation can 
be embedded within  the minimal SO(10) model, in a natural way. We show that none of the singlets of the Standard Model symmetries in the minimal set of
SO(10) representations can satisfy the conditions which are necessary for a scalar field to play the r\^ole of the inflaton. As a consequence, 
one has to introduce an extra scalar field, which however may spoil
the {\sl naturalness} of inflation within the context of SO(10).
Nevertheless, if we add an extra scalar field, we are then able to construct a model that can accommodate 
flat directions while it preserves the stability of the inflationary valley.}

\date{\today}

%\begin{document}
\maketitle
\flushbottom
%%%%%%%%%%%%%%%%%%%%%%%%%%%%%%%%%%%%%%%%%%%%%%%%%%%%%%%%%%%%%%%%%%%
\section{Introduction}\label{sec:intro}
Cosmological inflation is clearly the most studied and popular scenario that can provide an answer to some of the shortcomings that plague the hot big bang model, while it predicts a spectrum of adiabatic fluctuations that can fit the Cosmic Microwave Background (CMB) temperature anisotropies measurements~\cite{Ade:2013uln}.
Despite its success, one must however keep in mind that the inflationary scenario faces some problems, like the onset of inflation~\cite{Calzetta:1992gv,Calzetta:1992bp,Germani:2007rt} and the fine tuning of the parameters in the inflationary potential so that the inflationary predictions satisfy the data~\cite{Rocher:2004et,Rocher:2004my,Rocher:2006nh,Battye:2010hg}. Moreover, despite the more than three decades work in the subject, inflation still remains a paradigm in search of a model. 
If one accepts the validity of Grand Unified Theories (GUTs) and the standard thermal history of the universe, then one finds that the universe started at a symmetric phase with high temperature and then as the universe expanded and the temperature dropped, it underwent a series of Spontaneous Broken Symmetries (SSBs), followed by phase transitions (PTs), which could have left behind topological defects, as remnants of a previous more symmetric phase. Combining GUTs with Supersymmetry (SUSY), one can consider hybrid inflationary models, which may be of the F-term (often plagued by the $\eta$-problem, where contributions to the slow-roll parameters of the order of 1, due to Planck mass suppressed corrections to the inflaton potential, may impede a sufficiently long slow-roll period) or of the D-term type (leading always to cosmic string formation, due to the breaking of an extra U(1) symmetry)~\footnote{F-term inflation can be studied in the context of global supersymmetry, w!
 hereas D-term inflation must be addressed within supergravity~\cite{Rocher:2004et}.}.  
Given the plethora of precise data, arriving either from astrophysical  (in particular, the CMB), or from particle physics (in particular, the large Hadron Collider (LHC)) experiments, one can examine the validity of the various inflationary models and constrain their free parameters. Moreover, one can also study whether such models can arise naturally within the framework in which they have been proposed. Following this latter approach, we will study whether, within minimal supersymmetric SO(10), there is a singlet field that could play the r\^ole of the inflaton and thus realise an F-term hybrid supersymmetric inflationary model. 

In the first part of our study, we show that none of the (existing) scalar fields can satisfy the conditions necessary in order to play the r\^ole of the inflaton.  We thus introduce, in the second part of our analysis,  an extra SO(10)-singlet superfield and write down the most general Higgs superpotential. We can thus propose a 
model of F-term inflation embedded in SO(10) that can be in agreement with all current particle physics constraints.
Certainly F-term inflation can be realised within SO(10), but the necessity to introduce an extra singlet renders SO(10) less appealing as a gauge group describing the early evolution of our universe. We study the realisation of inflation within SO(10), because it is a well-studied gauge group in the context of hybrid inflation.

%%%%%%%%%%%%%%%%%%%%%%%%%%%%%%%%%%%%%%%%%%%%%%%%%%%%%%%%%%%%%%%%%%%%%
\section{Spontaneous symmetry breaking schemes within SO(10)}\label{sec:hypothesis}
The framework we will perform our analysis in is specified as follows:
\begin{itemize}
\item F-term hybrid inflation with superpotential~\cite{Dvali:1994ms},
\begin{equation}\label{eq:superpotentialFterm}
W^{\rm F}=\kappa S (\Psi\bar{\Psi}-M^2)~,
\end{equation}
where $S$ is a GUT singlet, $\bar\Psi$ and $\Psi$ are GUT Higgs fields in complex conjugate 
representations which lower the rank of the group by one unit when acquiring non-zero Vacuum Expectation
Values (VEVs), and $\kappa$ and  $M$ are two constants ($M$ has dimensions of mass) which can both be taken
positive with field redefinitions. 

The superpotential in Eq.~(\ref{eq:superpotentialFterm}) is the most
general superpotential consistent with an R-symmetry under which $W^F\rightarrow e^{i\beta}W^F$, $\bar\Psi\rightarrow e^{-i\beta}\bar\Psi$,
$\Psi\rightarrow e^{i\beta}\Psi$ and $S\rightarrow e^{i\beta} S$. The scalar potential has a valley of local minima for $S>S_{\rm crit}=M$, $\bar\Psi=\Psi$, and one global supersymmetric minimum at $S=0$, $\bar\Psi=\Psi=M$. 
Imposing initial conditions such that $S\gg S_{\rm crit}$, the fields quickly settle down the valley of local minima; the potential 
becomes $V=\kappa^2 M^4\neq 0$, supersymmetry is broken and inflation can take place. One-loop corrections to the effective scalar potential introduce a tilt and assist the scalar field $S$ to slowly roll down the valley of minima. When $S$ reaches a value below $S_{\rm crit}$, inflation stops by a waterfall regime and the fields settle down to the global minimum of the potential and supersymmetry gets restored.
\item A series of SSBs from SO(10)
  down to the Standard Model (SM) times Z$_2$, that does not generate
  harmful topological defects, like monopoles and domain walls, at the end of inflation. The discrete
  symmetry Z$_2$ must remain unbroken down to low energies, to ensure
  proton stability.  Following the detailed study presented in
  Ref.~\cite{Jeannerot:2003qv}, the SSB cascade should take one of
  the following  paths:
\begin{eqnarray}\label{eq:patternSSB}
{\rm SO}(10)&\rightarrow \dots \rightarrow {\rm
  G}_{3,2,2,{\rm B-L}}\rightarrow {\rm G}_{\rm SM}\times Z_2 \rightarrow {\rm SU}(3)_{\rm C} \times {\rm U}(1)_{\rm Q}
\times Z_2~, \\  \label{eq:patternSSB2} {\rm SO}(10)&\rightarrow \dots \rightarrow {\rm
  G}_{3,2,1,{\rm B-L}}\rightarrow {\rm G}_{\rm SM}\times Z_2 \rightarrow {\rm SU}(3)_{\rm C} \times {\rm U}(1)_{\rm Q}
\times Z_2~,
\end{eqnarray} 
where  ${\rm G}_{3,2,2,{\rm B-L}}$ and ${\rm
  G}_{3,2,1,{\rm B-L}}$ stand for ${\rm SU(3)}_{\rm C}\times {\rm
  SU(2)}_{\rm L}\times {\rm SU(2)}_{\rm R}\times {\rm U(1)}_{\rm B-L}$
and ${\rm SU(3)}_{\rm C}\times {\rm SU(2)}_{\rm L}\times {\rm
  U(1)}_{\rm R} \times {\rm U(1)}_{\rm B-L}$, respectively, and  $Z_2$ is the R-parity.
\item
Conservation of R-parity at low energies to accommodate proton
lifetime stability.  This requires the use of only ``safe'' Higgs
representations~\cite{Martin:1992mq}; thus one can use $\mathbf{10, 45, 54, 120,  
  126, 210}$ but not $\mathbf{16, 144, 560}$.
\item Only renormalisable contributions to the superpotential.
\item Type I or II see-saw mechanism. This requires a
$\mathbf{\overline{126}}_{\rm H}$ to participate to the Yukawa couplings to
fermions and appropriate Higgs couplings~\cite{Goh:2003hf}. The
type II may be more natural in the context of SO(10).
\end{itemize}
The above assumptions are compatible with the framework of
Ref.~\cite{Jeannerot:2003qv}, where the formation of cosmic strings
were found to be generic for a large number of SUSY GUTs.
To accommodate the CMB measurements one will then have to either fine
tune the parameters~\cite{Rocher:2004et,Rocher:2004my,Rocher:2006nh,Battye:2010hg}, 
or to complicate the models and render the strings unstable~\cite{Urrestilla:2004eh}.

Note that the GUT singlet $S$ in Eq.~(\ref{eq:superpotentialFterm}) needs not be a singlet of SO(10): in fact, inflation can be triggered at any stage in the SSB cascade that finally leads to the SM.
In the spirit of a minimal GUT SO(10), we will not add any SO(10) singlet, but rather look for the possibility that F-term hybrid inflation is triggered during the symmetry breaking cascade initiated by a minimal GUT Higgs field content.

%%%%%%%%%%%%%%%%%%%%%%%%%%%%%%%%%%%%%%%%%%%%%%%%%%%%%%%%%%%%%%%%%%%
\section{Inflation purely within minimal SO(10)}\label{sec:SO10}
Let us consider the following two well-studied classes of SO(10) models: the first one
is based on the Higgs content $\mathbf{210, 126, \overline{126},
10}$ \cite{Bajc:2004xe}; the second one focuses on realising a
doublet-triplet splitting and its Higgs content is
$\mathbf{54, 45, 45', 16, \overline{16}, 10, 10'}$~\cite{Chacko:1998jz},
sometimes extended by the introduction of singlets~\cite{Barr:1997hq}.

In the vein of the first class, it has been noticed that its minimal
Higgs content  is not fully able to account for the observed masses
and mixings of the fermions when the neutrino see-saw mechanism is
implemented~\cite{Aulakh:2005mw}. To cure this problem, it has been
proposed~\cite{Aulakh:2006hs} to enlarge the model with a $\mathbf{120}$.
In what follows, we will adopt this context to perform our study, following the principle of minimal number of Higgs fields.

%%%%%%%%%%%%%%%%%%%%%%%
\subsection{Higgs content and scalar superpotential}
%%%
The Higgs sector of the Lagrangian is based on the minimal model ({\sl
  see, e.g.}, Refs.~\cite{Bajc:2004xe,Fukuyama:2004xs,Aulakh:2005mw})
and contains the following superfields:
\begin{itemize}
\item $\Phi$ in the representation $\mathbf{210}$. In tensor notation,
  it is written as a fourth rank symmetric tensor $\Phi_{ijkl}$.
\item $\Sigma$ and $\bar{\Sigma}$ in the representations
  $\mathbf{126}$ and $\mathbf{\overline{126}}$, respectively. In
  tensor notation, they are written as an antisymmetric fifth rank tensor
  $\Sigma_{ijklm}$.
\item $H$ in the representation $\mathbf{10}$. In tensor notation, it is
written as  a 10-dimensional vector $H_i$.
\item $\Omega$ in the representation $\mathbf{120}$. In tensor
  notation, it is written as  a third rank anti-symmetric tensor
  $\Omega_{ijk}$.
\end{itemize}
Note that, all indices in the tensor notation are SO(10) indices and run from 1 to
10.  We will first examine whether inflation can be fully embedded
within this (minimal) field content, without introducing any
additional superfields.

Imposing that the superpotential is an SO(10) invariant with these
superfields, the most general Higgs superpotential can be written as
\begin{equation}\label{eq:superpotgeneralmin}
\begin{split}
\tilde{W}_{\rm H} = & m\, \Phi^2+\lambda \,\Phi^3 + m_H\, H^2 +m_\Sigma\, 
\Sigma \bar{\Sigma} +\eta\, \Phi \Sigma \bar{\Sigma}+\Phi H (\alpha\, 
\Sigma+\bar{\alpha}\, \bar{\Sigma})
\\
& + m_\Omega\,  \Omega^2+\beta\,  H \Phi \Omega+\gamma\, \Omega^2 \Phi+\Omega 
\Phi (\zeta\, \Sigma+\bar{\zeta}\, \bar{\Sigma})~.
\end{split}
\end{equation}
In the above expression, it should be understood that $\Phi^2$
means $\Tr\, \Phi^2=\Phi_{ijkl}\Phi_{ijkl}$, where a summation is
implicit on any repeated indices~\footnote{Using the notation with
  indices, it is necessary to understand why other contributions to
  the superpotential cannot exist, namely they would not be scalars of
  SO(10).}. The first line was obtained in Ref.~\cite{Bajc:2004xe}. The
second one adds all the terms that the new \textbf{120} Higgs
allows~\cite{Aulakh:2005mw}; this contribution is relevant only for
the precise fit of the SM fermion masses and will be omitted in the
following. Note that we have omitted contributions from the vector field
$H$, since it corresponds to the MSSM sector. 

One can easily notice that no term of the form of the second
contribution to Eq.~(\ref{eq:superpotentialFterm}) 
can be constructed with the current field content since $S$ would have
to be a singlet of SO(10) for the term $S M^2$ to be SO(10) invariant. However, as shown in
Ref.~\cite{Jeannerot:2003qv}, F-term inflation should not occur during
the first stage of SO(10) breaking, but at a later stage during the
SSB cascade, in order to solve the monopole problem. Thus, we shall
look for the F-term superpotential not in SO(10) notation but in ${\rm
  G}_{3,2,2,{\rm B-L}}$. Moreover, F-term inflation
should involve SM singlet components of the superfields of the theory
since their value at the end of inflation will not necessarily
vanish. Here we are not limiting ourselves to the SSB cascade via ${\rm G}_{3,2,2,{\rm B-L}}$:
 the other cascade via ${\rm G}_{3,2,1,{\rm B-L}}$ is also described in our formalism, as the fields that may play the r\^ole of the inflaton must have no charge under U(1)$_{\rm R}$.
Limiting ourselves to the fields satisfying the above requirements, the superpotential effectively reduces to~\cite{Bajc:2004xe}
\begin{equation}
\label{modified}
W_{\rm singlet}= m\, (p^2+3a^2+6b^2)+2\lambda\, (a^3
+3pb^2+6ab^2) +m_\Sigma\, \sigma \bar{\sigma}
+\eta\, \sigma \bar{\sigma}(p+3a-6b)~,
\end{equation}
where 
\begin{equation}
p=\Phi(1,1,1)~, \quad a=\Phi(15,1,1)~,
\quad b= \Phi(15,1,3)~,
\quad\sigma=\Sigma(\bar{10},1,3)~, \quad \bar{\sigma}=
\bar{\Sigma}(10,1,3)~, 
\nonumber
\end{equation}
and the integers in parenthesis indicate the representation under the Pati-Salam group ${\rm SU}(4)_{\rm C}\times {\rm SU(2)}_{\rm L} \times {\rm SU}(2)_{\rm R}$.
In terms of the other SSB cascade containing U(1)$_{\rm R}$, it is enough to replace the triplets with their ``neutral'' component.
Note that $\Omega(10,1,1)$ cannot be safely given a VEV
without breaking part of the Standard Model.
%~\footnote{Indeed, the
%  decomposition of the \textbf{10} of SU(4) under SU(3)$\times$ U(1)
% contains a component $\mathbf{1}(2)$ that is singlet under
%  SU(3)$_{\rm c}$ but charged under $\bl$. Since this component is
%  also a singlet under SU(2)$_{\rm R}$, it is not charged under
%  U(1)$_{\rm R}$. Thus, in the Pati-Salam model, the hypercharge of
%  the SM is written $Y/2=\pm I_{3R} +({\rm B-L})/2$, $\Omega(10,1,1)$ cannot
%  have a vanishing charge under $U(1)_Y$. The same argument can be
%  used to check that $\Sigma(\bar{10},1,3)$ can safely take a VEV.}.
  Furthermore, we do not consider $H$ as it has a small VEV because it contains the MSSM Higgs fields, and we do not expect it to play any significant r\^ole at the time of inflation.
A phase redefinition of the superfields can be used to set $m$, $m_\Sigma$ and $\eta$ to be real and positive, while $\lambda$ can be a complex couplings.

It is clear that the only two fields that belong to conjugate representations are $\sigma$ and $\bar{\sigma}$, therefore they must be the GUT Higgses that couple to the inflaton, like the superfields $\Psi$ and $\bar{\Psi}$ in Eq.(\ref{eq:superpotentialFterm}).
Let us address the question of whether we have an inflaton candidate. 
Clearly, $a$, $b$ and
$p$, though they all possess a coupling to $\sigma \bar{\sigma}$, they
also all have a mass term, which implies that none of them can play
the r\^ole of the inflaton field.  
We therefore want to find a combination of the fields $a$, $b$ and $p$ that couples with $\sigma \bar{\sigma}$ and is massless. Other conditions apply, but firstly we should ensure that there is a massless combination.

We will assume that the 3 superfields develop VEVs ($a_0$, $b_0$ and $p_0$), and they can be expanded around the new vacuum as:
\begin{equation}
p= p_0 + P~, \quad a= a_0 + A~,
\quad b= b_0 + B~.
\nonumber
\end{equation}
We also assume that the would-be inflaton is a linear combination of the three superfields, and not of their complex conjugates. 
We can then calculate the mass matrix for the complex scalar components of $\{A,B,P\}$ and, as a first step, check if there is a massless combination by imposing the vanishing of the determinant of the mass matrix.
This will give some relations between the vacua, and allow to define the  scalar field that could play the r\^ole of the inflaton.
In a second step, we will check whether other conditions necessary for inflation to start are satisfied.
In the mass matrix, we set to zero the VEVs of $\sigma$ and $\bar{\sigma}$, as it should be at the onset of inflation, and consider the VEVs to be complex.
The determinant, as a function of the VEVs of the fields, is
\beq
& \mbox{Det} \left(\mathcal{M}^2_{ij}\right) = \mbox{Det} \left( \sum_{k=A,B,P}\; \frac{\partial F_k}{\partial \phi_i} \frac{\partial F^*_k}{\partial \phi^*_j} \right) = \phantom{aaaaaaaaaaaaaaaaaaaaaaaaaaaaaaaaaaa} & \nonumber \\
& \phantom{aaaaaa} =   20736 \left| m^3 + (4a_0+p_0) \lambda m^2 + 2 (2a_0^2+a_0 p_0 - 7 b_0^2) \lambda^2 m - 12 a_0 b_0^2 \lambda^3 \right|^2~, & \label{eq:det}
\eeq
where $F_k = \partial W_{\rm singlet}/\partial \phi_k$ is the $F$-term associated with the scalar component $\phi_k$ of the superfield $k=A,B,P$, and the indices $i,j = A, B, P$ label the three relevant superfields.
In general, $\mbox{Det} \left(\mathcal{M}^2\right)$  is non-zero, unless some special conditions apply on the VEVs; we list below all  possible cases.

%%%%%%%%%%%%%%%%%%%%%%%%%%%%%%%%%%%%%%%%%%%%%%%
\subsubsection{Case $a_0\neq -\frac{m}{2 \lambda}$}
%%%%%%%%%%%%%%%%%%%
In this case, we can solve for $p_0$ and get
\beq
p_0 = \frac{12 a_0 b_0^2 \lambda^3 - 2 (2 a_0^2 - 7 b_0^2) \lambda^2 m - 4 a_0 \lambda m^2 - m^3}{m \lambda (m+2 \lambda a_0)}~. \label{eq:sol1}
\eeq
The massless eigenstate, candidate for the r\^ole of the inflaton, is given by
\beq
X =\frac{1}{N} \Big[-4 m \lambda b_0\; A + m (m+2 \lambda a_0) \; B -6 \lambda b_0 (m+2 \lambda a_0)\; P\Big]~, \label{eq:infl1}
\eeq
with $N$ a normalisation.
The other two combinations are massive, unless
\beq
|b_0|^2 = - \frac{ m^2 |m+2 \lambda a_0|^2}{4 |\lambda|^2 (13 m^2 + 36 m \Re( \lambda a_0) + 36 |\lambda a_0|^2)}\,, \label{eq:2zeromass}
\eeq
where we define the real part of $ \lambda a_0$ by $\Re(\lambda a_0)$.
However, it is easy to show that the above condition,  Eq.(\ref{eq:2zeromass}),  is never satisfied: in fact, the numerator is positive definite, so there exist a solution only if the denominator is negative, more precisely if $\Re(\lambda a_0)<0$  and
\beq
|\lambda a_0| > |\Re( \lambda a_0)| > \frac{36 |\lambda a_0|^2 + 13 m^2}{36 m} \Rightarrow \left( \frac{|\lambda a_0|}{m} \right)^2 - \frac{|\lambda a_0|}{m} + \frac{13}{36} < 0\,.
\eeq
The latter inequality is however never satisfied. Hence for $a_0 \neq - \frac{m}{2 \lambda}$, there can be only a single massless scalar, denoted by $X$ and given in Eq.~(\ref{eq:infl1}).

%%%%%%%%%%%%%%%%%%%%%%%%%%%%%%%%%%%%%%%%%%%%%%%%
\subsubsection*{Sub-case $b_0=0$ (and $a_0\neq -\frac{m}{2 \lambda}$)}
%%%%%%%%%%%%%%%%%%%%%%%%%%%%%
The previous case simplifies considerably for $b_0=0$. Imposing Eq.~(\ref{eq:sol1}), the mass matrix in the basis $\{A,B,P\}$ reads
\beq
\mathcal{M}^2 = \left( \begin{array}{ccc}
 36 |m+2\lambda a_0|^2 & 0 & 0 \\
0 & 0 & 0 \\
0 & 0 & 4 m^2
\end{array} \right).
\eeq
The inflaton candidate is therefore $B$ itself, while the other two fields $A$ and $P$ are always massive.

%%%%%%%%%%%%%%%%%%%%%%%%%%%%%%%%%%%%%%%%%%
\subsubsection{Case $a_0 = -\frac{m}{2 \lambda}$}
%%%%%%%%%%%%%%%%%%%%%%%%%%%%%%%%%
In this case one cannot solve for $p_0$, which disappears from Eq.(\ref{eq:det}), and the determinant reduces to
\beq
\mbox{Det} \left(\mathcal{M}^2\right) = 1327104\, m^2 |\lambda b_0|^4\,,
\eeq
therefore the presence of a massless mode requires $b_0=0$.
In this case, the mass matrix simplifies to
\beq
\mathcal{M}^2 = \left( \begin{array}{ccc}
0 & 0 & 0 \\
0 & 144 |\lambda p_0|^2  & 0 \\
0 & 0 & 4 m^2
\end{array} \right)
\eeq
and the massless field is $A$. There is a second massless field $B$ only if $p_0=0$.

%%%%%%%%%%%%%%%%%%%
\subsection{Conditions for inflation}
%%%%%%%%%%%%%%%%
We now study in detail the further conditions that ensure the existence of an inflationary potential, in order to pin down the successful VEV configurations.

%%%%%%%%%%%%%%%%%%%%%%%%%
\subsubsection{Case $a_0 = -\frac{m}{2 \lambda}$, $b_0 = 0$, $p_0 \neq 0$}
%%%%%%%%%%%%%
Let us start with the simple case $a_0 = -\frac{m}{2 \lambda}$, $b_0 = 0$ and $p_0 \neq 0$.
The massless field is $A$ and one can expand the superfields around the vacua, namely
\beq
a =  -\frac{m}{2 \lambda} + A\,, \qquad b = B\,, \qquad p = p_0 + P\,,
\eeq
to obtain the following superpotential for the superfields of the would-be inflaton $A$:
\beq
W_{\rm singlet} =  3 \left(\eta \sigma \bar{\sigma} -\frac{m^2}{2 \lambda} \right) A +  2 \lambda A^3 + 12 \lambda B^2 A + \mbox{other terms}\,.
\eeq
The first term in the superpotential is exactly of the form of Eq.(\ref{eq:superpotentialFterm}), however $A$ cannot play the r\^ole of the inflaton since  its superpotential contains a trilinear coupling.
Thus, this case is excluded.

%%%%%%%%%%%%%%%%%%%
\subsubsection{Case $a_0 = -\frac{m}{2 \lambda}$, $b_0 = p_0 = 0$}
%%%%%%%%%%%%%%%%%%%%%%%%%%%%%%%
In this case there are two massless fields, $A$ and $B$, hence the field that could play the r\^ole of the  inflaton must be a linear combination of these two fields.
Expanding around the VEVs, the superpotential for the massless scalars reads
\beq
W_{\rm singlet} = 3 \eta \sigma \bar{\sigma} (A-2 B) - 3 \frac{m^2}{2 \lambda} A+ 2 \lambda A (A^2 + 6 B^2) + 6 \lambda P  B^2 + \mbox{other terms}\,.
\eeq
This superpotential contains dangerous trilinear terms involving $A$ and $B$: in order to check the feasibility of this configuration of vacua, we can study the potential for the scalar components of the superfields, which contains
\beq
V_{\rm scalar} = 4 m^2 \phi_P^\ast \phi_P - 9 m^2 (\phi_A^2 + (\phi_A^\ast)^2) - 18 m^2 (\phi_B^2 + (\phi_B^\ast)^2) + \mbox{other terms}\,.
\eeq
The potential, therefore, contains mass terms for the real and imaginary parts of $A$ and $B$, and the mass has the wrong sign for the real parts.
This signals the fact that the vacuum configuration is not a local minimum of the scalar potential, therefore it cannot be used to trigger inflation.

%%%%%%%%%%%%%%%%%%%
\subsubsection{Case $a_0 \neq -\frac{m}{2 \lambda}$}
%%%%%%%%%%%%%%%%%%%%%%%%%%%%%%%%%
One can also in this case expand the superfields around the VEVs, as
\beq
a =  a_0 + A\,, \qquad b = b_0 + B\,, \qquad p = p_0 + P\,,
\nonumber
\eeq
where $p_0$ is related to $a_0$ and $b_0$ by Eq.(\ref{eq:sol1}), in order for the fields $A$, $B$ and $P$ to contain a massless eigenstate.
The field $X$ that could play the r\^ole of the inflaton is therefore given by Eq.(\ref{eq:infl1}), and we can express the fields $A$, $B$, and $P$ in terms of $X$, as
\beq
A = \frac{1}{N} \left( - 4 \lambda^* m b_0^\ast X + \dots \right)\ \ , \ \   B = \frac{1}{N} \left( m (m + 2 \lambda^\ast a^\ast_0 ) X + \dots \right)\ \ , \nn\\
  P = \frac{1}{N} \left( - 6 \lambda^\ast (m + 2 \lambda^\ast a^\ast_0) b^\ast_0 X + \dots \right)\,.
\eeq
to obtain a superpotential for $X$, which contains a trilinear term:
\beq
W_{\rm singlet} & \subset & 
 - \frac{4 m^2 (\lambda^\ast)^2 b^\ast_0}{N^3} \left[ 21 m^3 + 102 m^2 \lambda^\ast a^\ast_0 + 4 m (\lambda^\ast)^2 (39 (a^\ast_0)^2 + 8 (b^\ast_0)^2) + 72 (\lambda^\ast)^3 a_0^3\right] X^3  \nonumber \\
 & & + \mbox{interactions} + \mbox{other terms}\,.
\eeq
For the trilinear term to vanish, one should  impose a condition on the VEVs $a_0$ and $b_0$, namely
\beq
b_0 = 0\, \qquad\mbox{or}\, \quad b_0 = \pm i \frac{m+2 \lambda a_0}{4 \lambda} \sqrt{\frac{3 (7 m + 6 \lambda a_0)}{2 m}}\,. \label{eq:solb0}
\eeq
Let us now study these two cases in detail.

%%%%%%%%%%%%%%%%%%%%%%%%%%%%%%%%%%%%
\subsubsection*{Sub-case $a_0 \neq -\frac{m}{2 \lambda}$, $b_0 = 0$}
%%%%%%%%%%%%%%%%%%%%%%%%%%%%%%%%%%%%%%%%%%
In this case, the field that could play the r\^ole of the inflaton is $X = B$, with superpotential given by
\beq
W_{\rm singlet} = - 6 \eta \sigma \bar{\sigma} B + 6 \lambda (2 A + P)  B^2 + \mbox{other terms}\,.
\eeq
Since $B$ does not have a linear term, this case is excluded.

%%%%%%%%%%%%%%%%%%%%%%%%%%%%%%%%%%%%%%%%%%%%%
\subsubsection*{Sub-case $a_0 \neq -\frac{m}{2 \lambda}$, $b_0 \neq 0$}
%%%%%%%%%%%%%%%%%%%%%%%%%%%%%%%%%%%%%%%%%%%%%%%%%

Fixing the vacuum $b_0$ to the second solution in Eq.(\ref{eq:solb0}), the superpotential for $X$ contains both a linear term in $X$ and a coupling $\sigma \bar{\sigma} X$, as required, but also a dangerous quadratic term.
The quadratic term only vanishes when $a_0$ is real, i.e. $a_0 = a_0^*$, so that we will impose this condition from now on. 

The vanishing of the quadratic term is however still not enough to ensure that the would-be inflaton is massless: in fact, we assumed that the inflaton $X$ is a superposition of the chiral superfields.
The condition we imposed at the beginning, makes sure that a mass in the form $\phi_X^\ast \phi_X$ is zero, however it does not ensure the vanishing of mass terms in the form $(\phi_X^\ast)^2 + \phi_X^2$.
We numerically checked that there is no massless state once the full mass matrix, written in terms of real scalar fields, is considered in this vacuum structure.
We can therefore conclude that this last case is excluded.

Below we will therefore assume that the minimal
SO(10) is extended with the introduction of a singlet $S$ that will
play the r\^ole of the inflaton field.

%%%%%%%%%%%%%%%%%%%%%%%%%%%%%%%%%%%%%%%%%%%%%%%%%%%%%%%
\section{Extending the minimal SO(10)}\label{sec:extendedSO(10)}
%%%%%%%%%%%%%%%%%%%%%%%%%
Let us then introduce an extra scalar field $S$ that could play the r\^ole of the inflaton and examine whether we can find flat directions with a stable inflationary valley.
We will focus on the simple case where $S$ is a singlet of SO(10), while non-singlets may also be used to play the r\^ole of the inflaton~\cite{Antusch:2010va}.
%%%%%%%%%%%%%%%%%%%%%%%%%%%%%%%
\subsection{Higgs content and scalar superpotential}
%%%%%%%%%%%%%%%%%%%%%%%%%%%%%%%%%%%%%%%%%%%%%%%%%%%%%%%%%%%
The Higgs sector  is based on the minimal model
described in Refs.~\cite{Bajc:2004xe,Fukuyama:2004xs,Aulakh:2005mw},
with the additional introduction of an SO(10) singlet superfield $S$.
By imposing that the superpotential is a scalar function of the
superfields, the most general Higgs superpotential takes the form
\begin{equation}\label{eq:superpotgeneral}
\begin{split}
\tilde{W}_{\rm H} = & m\, \Phi^2+\lambda\,  \Phi^3 + m_H\,  H^2 +m_\Sigma \, 
\Sigma \bar{\Sigma} +\eta\, \Phi \Sigma \bar{\Sigma}+\Phi H (\alpha\,  
\Sigma+\bar{\alpha}\, \bar{\Sigma})\\
& + m_\Omega\, \Omega^2+\beta\,  H \Phi \Omega+\gamma\, \Omega^2 \Phi+\Omega 
\Phi (\zeta\,  \Sigma+\bar{\zeta}\, \bar{\Sigma})\\
& + \kappa\,  S(\Sigma \bar{\Sigma}-M^2)\\
& + m_S\, S^2 +\lambda_S\, S^3 + S(\delta_1\, H^2+\delta_2\, \Omega^2+
\delta_3\, \Phi^2)~.
\end{split}
\end{equation}
The first two lines contain the superpotential in Eq.~(\ref{eq:superpotgeneralmin}); the terms in the last two lines of the above expression appear because of the presence of an
extra singlet in the theory, included in order to realise inflation.

Of the above superpotential, we can safely neglect terms involving $\Omega$, because it does not contain a singlet component under the SM gauge symmetries, and $H$, since this superfield realises the electroweak
SSB and has therefore a very small VEV.
The third line in Eq.~(\ref{eq:superpotgeneral}) contains the superpotential terms required for F-term inflation.
The terms in the fourth line, containing the singlet field, are potentially dangerous as they can spoil hybrid inflation by generating mass or quartic terms for the inflaton field.
In the following, therefore, we will set all the extra terms containing $S$ to zero~\footnote{After carefully studying the general case, we found that an inflationary valley can also be found for tuned values of the extra couplings $m_{\rm S}$, $\lambda_{\rm S}$ and $\delta_3$. However, the minimum of the valley sits on a supersymmetric vacuum with vanishing scalar potential, therefore it cannot be used for hybrid inflation. One such solution is $\delta_3 = \lambda^2 \kappa M^2/(3 m^2)$, $\lambda_{\rm S} = - \delta_3^3/\lambda^2$ and $m_{\rm S} =- 3 m \delta_3^2/\lambda^2$.}: this shows that some tuning is necessary in order to obtain inflation.

The Higgs superpotential, relevant for our study, reads
\begin{equation}\label{eq:superpoteffectif}
%\begin{split}
{\tilde W}_{\rm H} = m\, \Phi^2+\lambda\,  \Phi^3 + m_\Sigma\,  \Sigma \bar{\Sigma} +\eta\,
\Phi \Sigma \bar{\Sigma} + \kappa\,
S(\Sigma \bar{\Sigma}-M^2)~.
%\end{split}
\end{equation}
Here, we can use the phases of the superfields to set $m$, $m_\Sigma$, $\kappa$ and $M$ to be real and positive, while $\lambda$ and $\eta$ may be complex couplings.
Following our results from the previous section, the inflaton must be contained in the SO(10) singlet $S$.

%%%%%%%%%%%%%%%%%%%%%%%%%%%%%%
\subsection{Vacuum expectation values and superfields}
We will follow the procedure of Ref.~\cite{Bajc:2004xe} to describe
how the cascade of SSB, given in Eq.~(\ref{eq:patternSSB2}), can be
realised. We need first to identify the components of the Higgs fields
that can take a non-vanishing VEV; they are necessarily singlets under
the SM. 
Using Ref.~\cite{Slansky:1981yr}, the only superfields that can be considered
are
\begin{equation}
\begin{split}
p=& \Phi(1,1,1)~, \quad a=\Phi(15,1,1)~,
\quad b=\Phi(15,1,3)~, \\
\sigma=&\Sigma(\bar{10},1,3)~, \quad \sigma=
\bar{\Sigma}(10,1,3)~, \quad s=S(1,1,1)~.
\end{split}
\end{equation}
This is the same set used in the previous section, with the addition of the SO(10) singlet.
The superpotential that one has to study reads
\begin{equation}
\begin{split}
W_{\rm H}= m\,(p^2+3a^2+&6b^2)+2\lambda\,(a^3
+3pb^2+6ab^2) +m_\Sigma\, \sigma \bar{\sigma}\\
&+\eta\, \sigma \bar{\sigma}(p+3a-6b)  +\kappa\, s(\sigma
\bar{\sigma}-M^2)~.
\end{split}
\end{equation}

%%%%%%%%%%%%%%%%%%%%%%%%%%%%%
\subsection{Minimisation of the superpotential}
%%%%%%%%%%%%%%%%%%%%%%%%%%%%%%%%%%%%
In the absence of a Fayet-Iliopoulos term $\xi$, as in our case, the
condition for the D-term, $(\xi+1/2\sum_iq_i \langle \Phi_i \rangle^2)^2$, to vanish is
\be 
\sum_iq_i \langle\Phi_i\rangle^2=0~,\nn
\ee
 where $q$ stands for the charge under
U(1) and $\langle\Phi_i\rangle$ denote the VEVs of the superfields in question. Since the only charged superfields are $\sigma$ and $\bar{\sigma}$,
which have opposite charges, the condition for the D-term to vanish is
$\langle \sigma \rangle=\pm \langle \bar{\sigma} \rangle$. The F-terms, $F_i \equiv \partial
  W_H/\partial \Phi_i$, read
\begin{equation}
\begin{split}
F_p & = 2m p +6
\lambda b^2 +\eta \sigma \bar{\sigma}~,\\
F_a &=3[2 m a +2 \lambda (2b^2+a^2) + \eta \sigma \bar{\sigma}]~,\\
F_b &=6[2 m b + 2 \lambda b (2a+p) - \eta \sigma \bar{\sigma}]~,\\
F_\sigma &=\bar{\sigma} [m_\Sigma +\eta (p +3a -6b) + \kappa s]~,\\
F_{\bar \sigma} &=\sigma [m_\Sigma +\eta (p +3a -6b) + \kappa s]~,\\
F_s &=\kappa(\sigma \bar{\sigma} -M^2) ~,
\end{split}
\end{equation}
and the scalar potential is the sum $V=\sum_i |F_i|^2$. The VEVs of the fields will take values in order to
minimise the scalar potential $V$.

%%%%%%%%%%%%%%%%%%%%%%%%%%%%%
\subsubsection{Global minima} \label{sec:globalmin}
%%%%%%%%%%%%%%%%%%%%%%%%%%%%%%

Let us study whether it is possible to choose the VEVs such that all
F-terms vanish, thus the potential itself vanishes, corresponding to
global (SUSY preserving) minima of the potential. 
Here we use a subscript $0$ to label the VEVs in order to distinguish them from the superfields, so that $s_0$ is the VEV of the superfield $s$, and so on.
The F-term
associated to $s$ vanishes only if $\sigma_0 = \bar{\sigma}_0 = \pm M$. To
construct a SUSY preserving global minimum, the other VEVs have to
satisfy the following conditions
\begin{equation}
\begin{split}
& 2m p_0 +6 \lambda b_0^2 +\eta M^2 =0~,\\
& 2 m a_0 +4 \lambda b_0^2 + \eta M^2 =0~,\\
& 2 m b_0 + 2 \lambda b_0 (2a_0+p_0) - \eta M^2=0~,\\
& m_\Sigma +\eta (p_0 +3a_0 -6b_0)+\kappa s_0 =0 ~.
\end{split}
\end{equation}
The latter equation sets the value of $s_0$:
\begin{equation}
s_0 = - \frac{m_\Sigma}{\kappa} + \frac{\eta}{\kappa} (6b_0 - p_0 -3a_0)\,.
\end{equation}
We first solve the system in the limit $\eta M^2\ll m^2$, where we can
approximate the equations by setting $M=0$. This approximation is a
realistic one, because in our setting $M^2$ is likely to be lower than
$m^2$, in order for the B-L symmetry breaking to occur as a second
stage in the SSB pattern, and for the validity of a perturbation
analysis, $\eta$ is required to be lower than 1. In this case, the
six solutions for the VEVs $a_0,b_0,p_0$ and $s_0$ are given by
\begin{itemize}
\item $p_0=0$, $a_0=0$, $b_0=0$, $s_0=-\frac{m_\Sigma}{\kappa}$;

\item $p_0=0$, $a_0=-\frac{m}{\lambda}$, $b_0=0$, $s_0=-\frac{m_\Sigma}{\kappa}+ 3 \frac{\eta m}{\kappa \lambda}$;

\item $p_0=-\frac{m}{3\lambda}$, $a_0=-\frac{m}{3\lambda}$,
$b_0=\pm\frac{m}{3\lambda}$, $s_0=-\frac{m_\Sigma}{\kappa}+ \frac{4 \pm 6}{3} \frac{\eta m}{\kappa \lambda}$;

\item $p_0=\frac{3m}{\lambda}$, $a_0=-\frac{2m}{\lambda}$,
$b_0=\pm\frac{im}{\lambda}$, $s_0=-\frac{m_\Sigma}{\kappa}+ 3(1 \pm 2 i) \frac{\eta m}{\kappa \lambda}$~.
\end{itemize}
When $M$ is not set equal to zero, there are six general solutions.
Two of them are found by noticing that $p_0=-b_0=a_0$ is a solution of the
system above. It gives
\begin{equation} 
\begin{split} 
p_0=-b_0~,\quad a_0=-b_0~, \quad b_0=\frac{m}{6 \lambda} \left( 1 \pm \sqrt{1-6\frac{\lambda \eta M^2}{m^2}}\right)~.
\end{split}
\end{equation}
The other four solutions are obtained by solving the following equation in $b_0$
\begin{equation}\label{eq:globalmini}
\begin{split}  
6\left( \frac{\lambda}{m} b_0 \right)^4 & + 2 \left(\frac{\lambda}{m} b_0  \right)^3 + \left(6 + \frac{\lambda \eta M^2}{m^2} \right) \left(\frac{\lambda}{m} b_0  \right)^2  \\
&  + 2 \left(1 + \frac{\lambda \eta M^2}{m^2} \right) \left( \frac{\lambda}{m} b_0 \right)  + \frac{\lambda \eta M^2}{m^2} =0~. 
\end{split}
\end{equation}
For these solutions, the VEVs $a_0$ and $p_0$ then read 
\begin{equation} 
p_0=- \frac{m}{\lambda} \left( \frac{1}{2} \frac{\lambda \eta M^2}{m^2} + 3 \frac{\lambda^2 b_0^2}{m^2} \right)~,\quad 
a_0=\frac{m}{\lambda} \left( \frac{3 \lambda^2 b_0^2}{2 m^2} - \frac{1}{2} + \frac{1}{4}  \frac{\lambda \eta M^2}{m^2} \left( 1 + \frac{m}{\lambda b_0} \right)\right)~.
\end{equation}
Let us now discuss the properties of these solutions. It is
interesting to note that the SO(10) preserving minimum found in
Ref.~\cite{Bajc:2004xe} (with $\sigma_0 = \bar{\sigma}_0 =p_0=a_0=b_0=0$) is not preserved by
our superpotential; SO(10) must be broken with the choice of
superpotential given in Eq.~(\ref{eq:superpoteffectif}). Note also
that if some SUSY preserving minima exist, they are never reached at a
vanishing VEV for the inflaton $s$, unless $m_\Sigma$ is tuned to vanish on the given solution.
The latter situation is however not generic, as the other terms in $s_0$ may have non-trivial phases, while $m_\Sigma$ is real and positive.
The symmetries preserved
by these minima are those of the SM, which is what is required at the
end of the B-L symmetry breaking.  

The exact equations determining the vacua depend on two combinations of parameters: $m/\lambda$ and $x =  \lambda \eta M^2/m^2$.
To better understand the vacua, we can expand the solutions for small $x \ll 1$ (which corresponds, at zero order, to the solutions for $M=0$): we will focus on 3 particular solutions that will be relevant for the onset of inflation.
At leading order in $x$, we find
\beq
& a_0 = \frac{m}{\lambda} \left( -1 + \frac{1}{4} x + \dots \right) \,, \quad b_0 = \frac{m}{\lambda} \left( - \frac{1}{2} x + \dots \right)\,, & \nonumber \\
& p_0 =  \frac{m}{\lambda} \left( - \frac{1}{2} x + \dots \right)\,, \quad s_0 = - \frac{m_\Sigma}{\kappa} + \frac{\eta m}{\kappa \lambda} \left( 3 - \frac{13}{4} x + \dots \right)\,; & \label{eq:min1}
\eeq
for the first solution, and
\beq
& a_0 = \frac{m}{\lambda} \left( -2 + \frac{3\pm i}{10} x + \dots \right) \,, \quad b_0 = \frac{m}{\lambda} \left( \pm i + \frac{1\pm 2 i}{10} x + \dots \right)\,, & \nonumber \\
& p_0 =  \frac{m}{\lambda} \left( 3 - \frac{2 \pm 9 i}{10} x + \dots \right)\,, \quad s_0 = - \frac{m_\Sigma}{\kappa} + \frac{\eta m}{\kappa \lambda} \left( 3 (1\pm2i) + \frac{4 \pm 3 i}{5} x + \dots \right)\,, & \label{eq:min2}
\eeq
for the remaining two.
From these approximate solutions we see that the vacua break SO(10) completely to the SM gauge symmetries.

%%%%%%%%%%%%%%%%%%%%%%%%%%%%%%%
\subsubsection{Local minima at the onset of inflation}
%%%%%%%%%%%%%%%
As a next step, one has to look for the local minimum of the potential
assuming an initially large value of the VEV of the inflaton $s$.
Indeed, this is the state of the field usually assumed in
chaotic inflation. To preserve the global picture of F-term
inflation, we will assume that the intermediate stage of symmetry
is obtained while being in the false vacuum corresponding to
$\sigma_0 =\bar{\sigma}_0  =0$, in order to minimise the contribution
from $F_\sigma$ and $F_{\bar{\sigma}}$ to the potential.

It is worth noting that contrary to the F-term inflation toy model 
 where only the large value of $s$ induces a mass term for the
$\sigma$ and $\bar{\sigma}$ fields, in our case here, also the mass term $m_\Sigma$ as well as the VEVs of
$p$, $a$ and $b$ make a contribution. Once the VEVs $\sigma_0$, and
$\bar{\sigma}_0$ vanish, the F-term VEVs read
\begin{equation}
\begin{split}
F_p & = 2m p_0 +6 \lambda b_0^2~,\\
F_a &=3[2 m a_0 +2 \lambda (2b_0^2+a_0^2)]~,\\
F_b &=6[2 m b_0 + 2 \lambda b (2a_0+p_0)]~,\\
F_\sigma  &=0 = F_{\bar \sigma}~,\\
F_s &= -M^2 \kappa ~.
\end{split}
\end{equation}
As a consequence, even if the first three F-terms in the scalar potential $V$ can be cancelled
by an appropriate choice of $p_0$, $a_0$ and $b_0$, the potential will
be constant and given by $V_0=\kappa^2 M^4$; this is an $s-$flat direction. The 
minimum of the potential is obtained for the VEVs of $a_0$, $b_0$ and $p_0$ that set to zero the associated $F$-terms. 
The six solutions are the same as the ones we found for the global minima for $M=0$:
\begin{itemize}
\item $p_0=0$, $a_0=0$, $b_0=0$~. This minimum is obviously invariant 
under SO(10).

\item $p_0=0$, $a_0=-\frac{m}{\lambda}$, $b_0=0$~. Since $a_0\equiv \langle
  \Phi(15,1,1)\rangle$, it is clear that SU(2)$_{\rm L}\times$ SU(2)$_{\rm R}$ is
  preserved by this minimum. The component of the \textbf{15} of
  SU(4)$_{\rm C} \supset {\rm SU(3)}_{\rm C}\times \bl$ that can take
  a VEV is the one that preserves SU(3)$_{\rm C}$, which is also
  uncharged under $\bl$~\cite{Slansky:1981yr}. This minimum is thus
  invariant under G$_{3,2,2,{\rm B-L}}$.

\item $p_0=\frac{3m}{\lambda}$, $a_0=-\frac{2m}{\lambda}$,
  $b_0=\pm\frac{im}{\lambda}$~. For the symmetries left unbroken by these
  minima, this case is similar to the above one (since the VEV $p_0$
  has no effect on symmetries), except that the VEV $b_0\equiv \langle
  \Phi(15,1,3)\rangle$ induces the additional breaking SU(2)$_{\rm R}
  \rightarrow {\rm U(1)}_{\rm R}$. This minimum is thus invariant under
  G$_{3,2,1,{\rm B-L}}$.

\item $p_0=-\frac{m}{3\lambda}$, $a_0=-\frac{m}{3\lambda}$,
  $b_0=\pm\frac{m}{3\lambda}$~. A careful analysis of these minima shows
  that they are invariant under SU(5)$\times {\rm U(1)}$~\cite{Bajc:2004xe}.

\end{itemize}
Note that we have not made any assumption on the values of the
potential parameters and our solutions are exact. These
solutions, already found in Ref.~\cite{Bajc:2004xe}, are now of phenomenological
interest even though they
do not give rise to the SM, since inflation will drive the last part of the symmetry
breaking.

%%%%%%%%%%%%%%%%%%%%%%%%%%%%%%%%%%%
\subsubsection{Stability of the inflationary valley}
%%%%%%%%%%%%%%%%%%%%%%

The vacua that we are interested in are the ones that have an unbroken 
G$_{3,2,2,{\rm B-L}}$ and G$_{3,2,1,{\rm B-L}}$ symmetry. To compute the
scalar potential for the former case ($a_0 = - m/\lambda$ and $b_0 = p_0 = 0$), we expand the scalar components of the superfields around the local vacua
\begin{equation}
a=\varphi_a-\frac{m}{\lambda}, \quad b=\varphi_b, \quad p=\varphi_p, \quad \sigma = \varphi_\sigma, \quad \bar{\sigma} = \varphi_{\bar{\sigma}},  \quad s = s_0 + \varphi_s \,,
\end{equation}
where $\varphi_x$ are the scalar perturbations around the vacuum expectation
value of the field $x$. The scalar potential then reads
\begin{equation}
\begin{split}
V& =\Big|2m\varphi_p+6\lambda\varphi_b^2+ \eta \varphi_\sigma \varphi_{\bar{\sigma}}\Big|^2+\Big| 12m \varphi_b + 2 \lambda \big( 6 \varphi_p \varphi_b+12(\varphi_a-\frac{m}{\lambda}) \varphi_b \big)-6 \eta \varphi_\sigma \varphi_{\bar{\sigma}} \Big|^2\\ &+\Big| 6m( \varphi_a - \frac{m}{\lambda})+ 2 \lambda\big( 3 (\varphi_a - \frac{m}{\lambda})^2+6\varphi_b^2\big)+3\eta\varphi_\sigma\varphi_{\bar{\sigma}}\Big|^2+\kappa^2\Big|\varphi_\sigma\varphi_{\bar{\sigma}}-M^2\Big|^2\\ &+(|\varphi_\sigma|^2+|\varphi_{\bar{\sigma}}|^2) \Big| m_\Sigma + \eta (\varphi_p + 3\varphi_a - 3 \frac{m}{\lambda} - 6\varphi_b)+\kappa \big( s_0 + \varphi_s \big) \Big|^2~.
\end{split}\label{stability1}
\end{equation}
Expanding the potential up to quadratic terms, we have
\begin{equation}
\begin{split}
V& = \kappa^2 M^4 + 36 m^2 \varphi_a^* \varphi_a + 144 m^2 \varphi_b^* \varphi_b + 4 m^2 \varphi_p^* \varphi_p + \\ & \Big| m_\Sigma - \frac{3 \eta m}{\lambda} + \kappa s_0 \Big|^2 (\varphi_\sigma^* \varphi_\sigma + \varphi_{\bar{\sigma}}^* \varphi_{\bar{\sigma}}) - \kappa^2 M^2 (\varphi_\sigma \varphi_{\bar{\sigma}} + \varphi_\sigma^* \varphi^*_{\bar{\sigma}}) + \dots
\end{split}
\end{equation}
We note that the scalar perturbations in $a$, $b$ and $p$ correspond to massive fields, while the scalar perturbations in $\sigma$ and $\bar{\sigma}$ have a mass matrix that depends on the VEV of the inflaton.
For large values of $s_0$, the mass squares are positive and the vacuum $\sigma_0 = \bar{\sigma}_0 = 0$ is stable.
During inflation, the value of $s_0$ will slowly roll along the flat direction, until the condition
%This vacuum is stable, i.e. it has a flat direction for $\varphi_\sigma = \varphi_{\bar{\sigma}}^*$, for
\begin{equation}
s_0^{\rm crit} = - \frac{m_\Sigma}{\kappa} + 3\frac{\eta
  m}{\kappa \lambda} \pm M\label{stabilitycond}
\end{equation}
is met: this is the critical value of the inflaton VEV below which the $\varphi_\sigma$--$\varphi_{\bar{\sigma}}$ system will develop a tachyonic mass and the system will quickly settle on a stable vacuum.
For small $M$, the unstable point is close to the minimum in Eq.~(\ref{eq:min1}), so it is likely that the fields will settle on this minimum: at this point, supersymmetry is restored and SU(2)$_{\rm R} \times$ U(1)$_{\rm B-L}$ is broken to the hypercharge by the non-vanishing vacua of $\sigma$ and $\bar{\sigma}$ at a scale $M$.

We can now repeat the calculation in the latter case, which corresponds to the ${\rm G}_{3,2,1,{\rm B-L}}$ SSB cascade, by expanding the fields around the VEVs as follows:
\begin{equation}
a=\varphi_a - 2 \frac{m}{\lambda}, \quad b= \pm i \left( \varphi_b + \frac{m}{\lambda} \right), \quad p=\varphi_p + 3 \frac{m}{\lambda}, \quad \sigma = \varphi_\sigma, \quad \bar{\sigma} = \varphi_{\bar{\sigma}},  \quad s = s_0 + \varphi_s \,.
\end{equation}
Following the same procedure as for the former case, the potential up to quadratic terms in the fields is given by
\begin{equation}
\begin{split}
V& = \kappa^2 M^4 + 900 m^2 \varphi_a^* \varphi_a + 720 m^2 \varphi_b^* \varphi_b + 148 m^2 \varphi_p^* \varphi_p \\ 
& + 432 m^2 (\varphi_a^* \varphi_b + \varphi_b^* \varphi_a) + 288 m^2 (\varphi_a^* \varphi_p + \varphi_p^* \varphi_a) - 24 (\varphi_p^* \varphi_b + \varphi_b^* \varphi_p) \\
& \Big| m_\Sigma - \frac{3 (1\pm 2 i) \eta m}{\lambda} + \kappa s_0 \Big|^2 (\varphi_\sigma^* \varphi_\sigma + \varphi_{\bar{\sigma}}^* \varphi_{\bar{\sigma}}) - \kappa^2 M^2 (\varphi_\sigma \varphi_{\bar{\sigma}} + \varphi_\sigma^* \varphi^*_{\bar{\sigma}}) + \dots
\end{split}
\end{equation}
Once more, the inflaton field is massless, and the first two lines define a mass matrix for the three complex fields $\varphi_{a,b,p}$, whose eigenvalues are numerically given by $m_i = \{ 1.841, 21.768, 35.927\} \cdot m$.
The valley is stable, until the inflaton VEV reaches the critical value
\begin{equation}
s_0^{\rm crit} = - \frac{m_\Sigma}{\kappa} + \frac{3 (1\pm 2 i) \eta m}{\kappa \lambda} \pm M.\label{stabilitycond2}
\end{equation}
For small $M$, the nearest global vacuum is one of the two in Eq.~(\ref{eq:min2}), which also restore supersymmetry and break the remaining symmetries down to the SM ones.

We were thus able to propose a model and a superpotential such that
F-term inflation is explicitly embedded in a detailed and minimal model of
SO(10) that has successfully passed particle physics phenomenology
tests. We have found three local minima (out of six) for the scalar potential
for which the symmetries are such that no harmful topological defects
are formed at the end of inflation and where there are no tachyonic
modes that will destabilise the inflationary valley:
\begin{equation}\label{eq:localmini1}
\sigma_0 = \bar{\sigma}_0=p_0=b_0=0~,\quad a_0=-\frac{m}{\lambda}~,\quad s_0\neq 0~,  \quad
V_0=\kappa^2 M^4~,
\end{equation}
or
\begin{equation}\label{eq:localmini2}
\sigma_0 = \bar{\sigma}_0=0~,\quad a_0=-2\frac{m}{\lambda}~, 
\quad p_0=3\frac{m}{\lambda}~, \quad b_0=\pm i\frac{m}{\lambda}~, \quad s_0\neq 0~, 
\quad V_0=\kappa^2 M^4~.
\end{equation}
The first local minimum can give rise to a successful phase of F-term
inflation that will dynamically break G$_{3,2,2,{\rm B-L}}$ into the G$_{\rm SM}\times Z_2$ symmetry group, thus
realising the SSB patterns of Eq.~(\ref{eq:patternSSB}).  
The latter two minima will do the same with G$_{3,2,1,{\rm B-L}}$.
Cosmic
strings are formed at the end of inflation~\cite{Jeannerot:2003qv} at an
energy scale related to inflationary physics and are expected to have
some impact in cosmology~\cite{Rocher:2004et,Rocher:2004my,Ade:2013xla}.  By
doing so, the system will evolve to one of the minima detailed in Section~\ref{sec:globalmin}.

%%%%%%%%%%%%%%%%%%%%%%%%%%%%%%%%%%%%%%%%%%
\section{Conclusions}
%%%%%%%%%%%%%%%%%%%%%%%%%%%%%%%%%%%%%%%%%%%%
The inflationary paradigm has been extensively studied in the context of Supersymmetric Grand Unified Theories.
Given that SO(10) is a well-studied gauge group, we have investigated whether it can accommodate an inflationary era
without the introduction of an extra scalar field to play the r\^ole of the inflaton. In particular, we have studied
whether F-term hybrid inflation can be incorporated in a rather natural way. We have shown that none of the scalar fields of SO(10) can play the r\^ole of the inflaton and one has to introduce an extra scalar field. This result may be considered as an element that spoils the naturalness of inflation within SO(10). 

Adding an extra scalar field, singlet under SO(10), that could play the r\^ole of the inflaton, we have shown the existence of an appropriate superpotential that can have flat directions preserving the stability of the inflationary valley.

%%%%%%%%%%%%%%%%%%%%%%%%%%%%%%%%%%%%%%%%%
\begin{acknowledgments}
It is a pleasure to thank Jonathan Rocher for his collaboration on the early stages of this work.
\end{acknowledgments}

\end{document}